\documentclass[aps,prb,showpacs,groupedaddress,
twocolumn]
{revtex4}
\usepackage{graphicx,bm}

\begin{document}

\title{Increase of critical currents and peak effect
in Mo substituted YBa$_{2}$Cu$_{3}$O$_{7}$}

\author{K. Rogacki}
\affiliation {Institute of Low Temperature and Structure Research,
Polish Academy of Sciences, 50-950 Wroclaw, P.O.Box 1410, Poland}
\affiliation {Laboratory for Solid State Physics, ETH Z\"{u}rich,
8093 Z\"{u}rich, Switzerland}

\author{B. Dabrowski and O. Chmaissem}
\affiliation {Physics Department, Northern Illinois University,
DeKalb, IL 60115}

\date{\today}

\begin{abstract}
Superconducting critical currents $j_{c} > 10^{5}$ A/cm$^{2}$ at
temperatures $T \sim 50$~K and magnetic fields $B \sim 6$~T are
reported for the YBa$_{2}$Cu$_{3-x}$Mo$_{x}$O$_{7+d}$ compound
with $x = 0.02$. Clear evidence for the increased pinning force
$F_p$ was found from a peak effect present for $j_{c}(B)$. The
pinning force was analyzed by a scaling procedure using Kramer's
approach. For a wide range of fields and temperatures, we were
able to express all $F_p$ data as a single function of a reduced
field $b = B/B_k$, where the scaling field $B_{k} << H_{c2}$ was
related to the irreversibility filed $B_{irr}$. Analyses of the
field dependence of $j_{c}(B,T)$ and $F_{p}(b)$ show that the
effective pinning centers act as weakly interacting extended
point-like defects. We propose that the pinning centers are
randomly distributed small defects, most likely the dimers of
MoO$_6$ octahedra in the CuO chains.
\end{abstract}



\maketitle

\section{INTRODUCTION}

One of the most promising areas where high temperature
superconductors (HTSC) can find immediate utilization is
large-scale power application. Such applications require
development of materials capable of sustaining critical currents
on the order of $j_{c} > 10^5$ A/cm$^{2}$ at liquid nitrogen
temperatures and displaying only a weak dependence on applied
magnetic fields. These critical current requirements are difficult
to satisfy in ceramic HTSC materials. However, increased $j_c$
values and weaker $j_c$ dependencies on applied field can be
achieved by increasing the flux pinning, which depends on the
interaction of magnetic vortices with crystal defects, and with
second-phase impurities including magnetic
nanoparticles.\cite{Jooss'PRL'1999,Martinez'PRB'1996,Rogacki'PRB'2000,
Haugan'Nature'2004,MacManus'NatureMat'2004,Shlyk'APL'2005,Zhao'SST'2005,
Snezhko'PRB'2005} One of the most effective methods of increasing
the flux pinning at intermediate range of applied fields is
creation of columnar irradiation
defects.\cite{Civale'SST'1997,Niebieskikwiat'PRB'2001,
Thompson'PhysicaC'2002} Another effective method is introduction
of microscopic grains of the Y$_2$BaCuO$_5$ phase that are the
most efficient at the intermediate range of temperatures.
\cite{Martinez'PRB'1996,Haugan'Nature'2004} Both these methods are
rather difficult to implement in large-scale applications. Thus,
simpler methods are required for intensifying flux pinning in the
range of increased temperatures and high magnetic fields. We show
here that optimized chemical substitution could be one such
methods for the YBa$_2$Cu$_3$O$_7$ type superconductors.

In YBa$_2$Cu$_3$O$_7$ (Y123) several cation substitutions were
made for the small ion (copper) and for the large ions (barium and
yttrium).\cite{SeeRogacki'PRB'2000,Tarascon'PRB'1988,Veal'APL'1989}
In most cases, substitutions for copper were found to rapidly
decrease $T_c$. For the Fe and Al substitutions on the Cu-chain
site, $T_c$ remained constant up to 5 and 10~\%,
respectively.\cite{Tarascon'PRB'1988} For Ga, $T_c$ decreased more
rapidly resulting in the complete absence of superconductivity at
10~\% doping.\cite{Lin'PhysicaC'1996} A modest increase of $T_c$
was conceivably observed by substitution of 2-5~\%
Co.\cite{Tarascon'PRB'1988} For Sr substituted
YBa$_{2-y}$Sr$_y$Cu$_3$O$_7$, a gradual decrease of $T_c$ was
observed up to the solubility limit found around $y = 1$ under
normal synthesis conditions.\cite{Veal'APL'1989} Interestingly,
when the high oxygen pressure anneal (HPA) was applied for the
Mo-substituted YBaSrCu$_{3-x}$Mo$_x$O$_{7+d}$ compound, $T_c$
increased from 81 to 87~K.\cite{Rogacki'PRB'2000} However,
decrease of $T_c$ and $j_c$ has been reported for the
Mo-substituted Y123 compounds synthesized under ambient
pressure.\cite{Kuberkar'ApplSup'1995} Here we show that both the
higher $T_c$ and increased $j_c$ can be achieved for the
YBa$_2$Cu$_{3-x}$Mo$_x$O$_{7+d}$ system if appropriate synthesis
and oxygenation conditions are applied.

In this work, an increase of the superconducting $T_c$, intragrain
persistent critical current density $j_c$, and irreversibility
field $B_{irr}$, was achieved for the
YBa$_2$Cu$_{3-x}$Mo$_x$O$_{7+d}$ compound by determining the
optimum composition, synthesis, and annealing conditions. The
structural and superconducting properties of the optimized
materials with $T_c$ $\approx 90$-93~K are examined and compared
with recently published data for similar compounds. The
magnetization $M(B)$ isotherms are measured over wide field and
temperature ranges to derive $j_{c}(B)$ and $j_{c}(T)$ which
reveal a pronounced peak effect. A scaling approach is used to
determine the mechanism responsible for the peak effect.  We show
that substitution of Mo results in creation of additional pinning
centers and formation of the significant peak effect at elevated
fields and temperatures. We propose that the small-size defects in
the CuO chains, most likely the Mo$_2$O$_{11}$ dimers made of
nearest neighbor MoO$_6$ octahedra, may act as effective pinning
centers similar to that seen in YBaSrCu$_{3-x}$Mo$_x$O$_{7+d}$
compounds.\cite{Rogacki'PRB'2000} The Mo$_2$O$_{11}$ dimers may
locally perturb superconductivity in the CuO$_2$ planes by
providing additional weakly interacting pinning centers which
increase the total pinning force. Finally, we show an important
difference between the pinning mechanisms observed for the pure
Y123 ($x = 0$) and the Mo-substituted ($0 < x \leq 0.05$) samples
for which substitution results in the pronounced maximum in
$j_{c}(B)$ at high fields.

Recently, intriguing clustering of Al ions substituted for Cu in
the Cu-O chains of YBa$_2$Cu$_3$O$_7$ has been
reported.\cite{Li'PRB'2004} This clustering, studied by positron
annihilation and x-ray diffraction experiments, induces the
localization of hole carriers, weakens the function of the Cu-O
chains as carrier reservoir, and as a consequence, suppresses
superconductivity. If the Al clusters are separated one from
another by at least few unit cells, they should disturb
superconductivity only locally. A similar effect of small
Mo$_2$O$_{11}$ defects in the Cu-O chains separated by several
unit cells is considered in this work as a possible mechanism
responsible for creation of pinning centers in the Mo substituted
YBa$_2$Cu$_3$O$_7$ samples.

\section{EXPERIMENTAL DETAILS}

\begin{figure}[!htb]
\includegraphics*[width=0.43\textwidth]{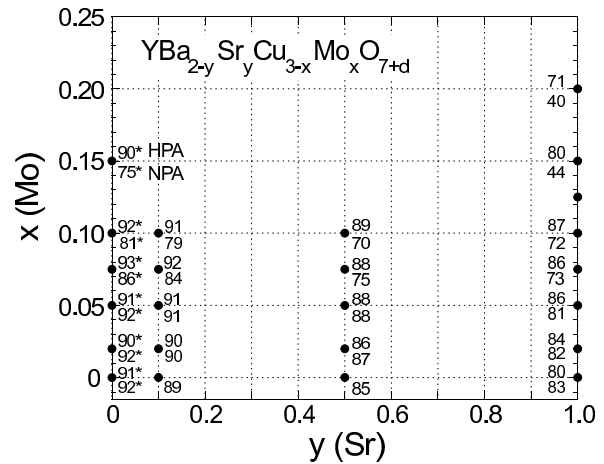}
\caption{Superconducting transition temperatures, $T_c$, for the
YBa$_{2-y}$Sr$_y$Cu$_{3-x}$Mo$_x$O$_{7+d}$ samples annealed in
oxygen at normal pressure (NPA, lower numbers) and high pressure
(HPA, upper numbers). Stars mark $T_c$'s for samples investigated
in this work.} \label{x(Mo)y(Sr)}
\end{figure}

Polycrystalline samples of YBa$_2$Cu$_{3-x}$Mo$_x$O$_{7+d}$ with
$0 \leq x \leq 0.2$ were synthesized from a mixture of Y, Mo, and
Cu oxides, and Ba carbonate. These samples compose a subset of a
wider range of recently investigated materials
YBa$_{2-y}$Sr$_y$Cu$_{3-x}$Mo$_x$O$_{7+d}$ shown in
Fig.~\ref{x(Mo)y(Sr)}. Samples were fired in air and oxygen at
880-920~$^{o}$C for several days with frequent intermediate
grindings.  Normal pressure anneals (NPA) were done in pure oxygen
for 12~h at 600~$^{o}$C followed by slow cooling to room
temperature. High pressure anneals (HPA) were done in pure oxygen
using 250-300 bar at 650~$^{o}$C followed by slow cooling to room
temperature. Sample homogeneity was checked by powder x-ray
diffraction on a Rigaku D/MAX Diffractometer using Cu$_{K\alpha}$
radiation. Structural parameters and oxygen contents were refined
from time-of-flight neutron powder diffraction data collected on
the General Purpose Powder Diffractometer (GPPD) at the Argonne
National Laboratory Intense Pulsed Neutron Source (IPNS).
Diffraction data for the NPA and HPA samples were collected at
room temperature. High-resolution backscattering data were
analyzed using the Rietveld method with the General Structure
Analysis System (GSAS) code.\cite{Larson'LAUR'2000} In the
analysis, background, absorption, peak width, and extinction
parameters were refined, together with the lattice parameters,
atomic positions, and isotropic temperature factors for the
cations and oxygen atoms. The oxygen contents were also determined
by thermogravimetric analysis using a Cahn TG171 system with slow
(0.6~deg/min) heating and cooling rates. The sizes and shapes of
the grains were examined using a Hitachi Scanning Electron
Microscope (SEM).

\begin{figure}[!htb]
\includegraphics*[width=0.37\textwidth]{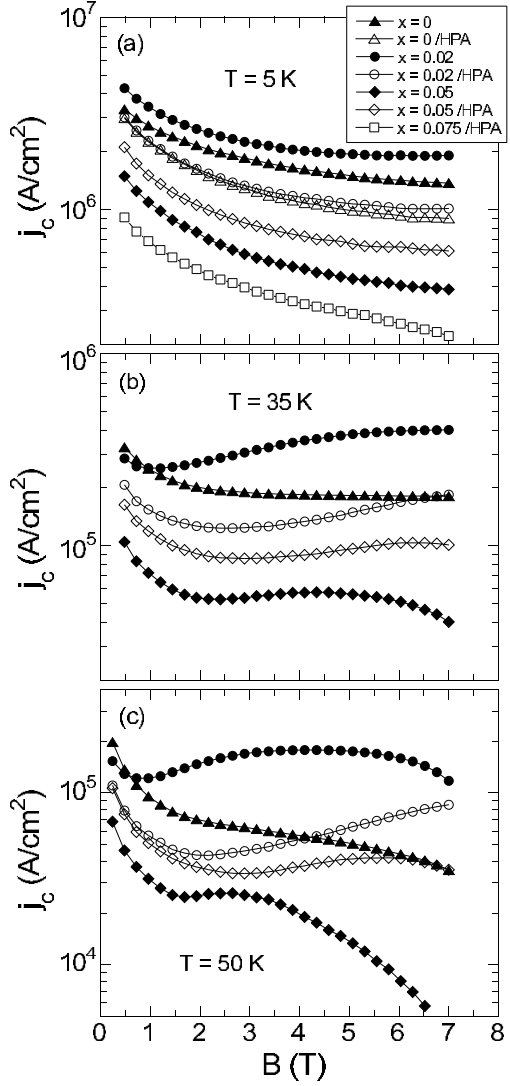}
\caption{Persistent critical currents, $j_c$, versus magnetic
field at 5, 35, and 50 K for the YBa$_2$Cu$_{3-x}$Mo$_x$O$_{7+d}$
samples with $x =$ 0 (triangles), 0.02 (circles), 0.05 (diamonds),
and 0.075 (squares) annealed in oxygen at normal pressure (closed
symbols) and high pressure (open symbols, HPA). An extended peak
effect (fishtail for magnetization loops) is observed for the
substituted compounds only.} \label{j(B)5,35,50K}
\end{figure}

Resistivity, $R$, ac susceptibility, $\chi$, and dc magnetization,
$M$, were measured with a Quantum Design Physical Properties
Measurement System (PPMS) equipped with 7~T superconducting
magnet. For resistivity measurements, solid pieces of the sintered
samples were used and the standard four-lead dc method was applied
with a current density of about 0.2~A/cm$^2$. Solid pieces were
also used for $\chi$ measurements to estimate the superconducting
shielding effect. Powder samples with masses of about 200~mg were
employed for both $\chi$ and $M$ measurements to obtain the
superconducting volume fraction, critical currents, and
irreversibility fields (Figures \ref{M(B)5K} to \ref{Fp(B)}). The
susceptibility was taken upon warming from the zero-field-cooled
(ZFC) state using the ac field of 1~Oe at 200~Hz. For these
measurements, the real $\chi'$ and imaginary $\chi''$ parts of the
signal were collected. As will be discussed in following sections,
$\chi''$ was used to determine the irreversibility field
$B_{irr}$. This is based on a principle that when the energy
losses are of hysteretic type, the energy absorption $\chi''$ is
proportional to the area of the magnetization hysteresis loop
which is non-zero in the presence of irreversibility of $M(T)$.

\section{SYNTHESIS AND INITIAL CHARACTERIZATION}

The optimum synthesis and annealing conditions of temperature and
oxygen pressure were determined to maximize the superconducting
transition temperature $T_c$.  Samples were fired several times in
air, 1~$\%$ oxygen in Ar, and pure oxygen at increasing
temperatures, checked for phase purity, annealed under several
oxygen pressure and temperatures, and checked for $T_c$. All
compositions were single phase when synthesized in air or oxygen
at 880-920~$^{o}$C.  Small amounts of impurities $(\approx 3~\%)$
were present for synthesis temperatures both lower than
880~$^{o}$C, due to incomplete reaction, and higher than
920~$^{o}$C, due to partial melting. High-pressure oxygen anneals
at 250-300 bar and at temperatures around 650~$^{o}$C followed by
slow cooling (0.2 deg/min) were found to be optimal to increase
the oxygen content and to attain the highest $T_c$ for samples
obtained from synthesis in pure oxygen.  After the same HPA, the
$T_c$'s of the samples synthesized in air and 1~$\%$ oxygen were
consistently lower by several degrees. Attempts to synthesize the
YBa$_2$Cu$_{3-x}$Mo$_x$O$_{7+d}$ compounds directly at high oxygen
pressure of 600 bar at 950~$^{o}$C produced multiphase materials
with $T_c$'s $\sim$ 60~K.

Figure~\ref{x(Mo)y(Sr)} shows the maximum $T_c$'s we have obtained
for the YBa$_{2-y}$Sr$_y$Cu$_{3-x}$Mo$_x$O$_{7+d}$ samples with $0
\leq x \leq 0.2$ and $0 \leq y \leq 1$. The upper and lower
numbers denote the $T_c$'s obtained after anneal in oxygen and
high pressure, respectively. The oxygen index, $d$, is clearly
positive for the Mo-substituted compositions (e.g., for the $y =
1$ samples, $d = 0.16$ and 0.3 for $x = 0.1$ and 0.2,
respectively).\cite{Rogacki'PRB'2000} The $T_c$ for the NPA $y =
0$ samples decreases rapidly with $x$ from 92~K for $x = 0$ to
81~K for YBa$_2$Cu$_{2.9}$Mo$_{0.1}$O$_{7.14}$. However, the HPA
restores $T_c \approx 92$~K for all $0 < x \leq 0.1$ samples. The
main effect of the HPA is to fill the oxygen vacancies in the CuO
chains (i.e., to increase the hole-carrier concentration) and to
introduce the inter-chain oxygen atoms (i.e., to form defects).
These electronic and structural changes cause important
modifications of the superconducting properties that will be
discussed in subsequent sections.

\begin{figure}[!htb]
\includegraphics*[width=0.35\textwidth]{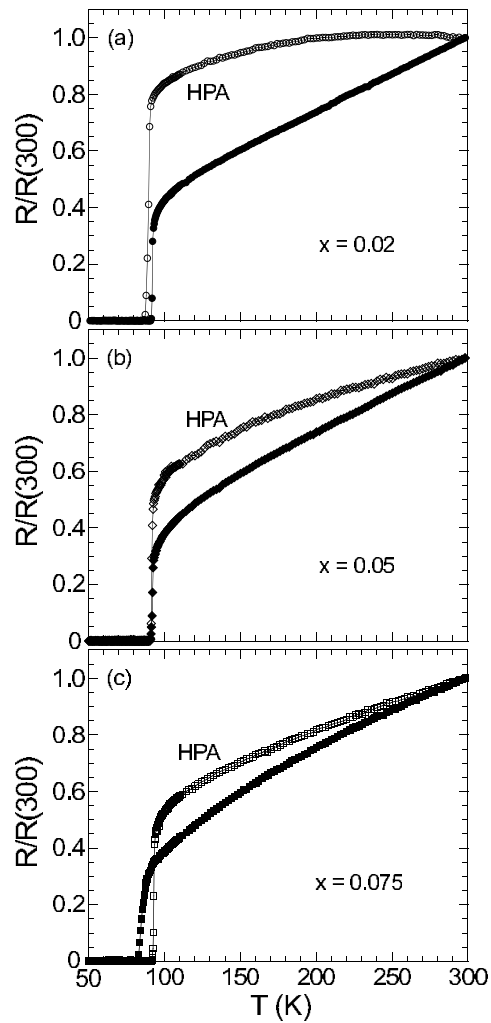}
\caption{Normalized resistance, $R/R(300)$, versus temperature for
the YBa$_2$Cu$_{3-x}$Mo$_x$O$_{7+d}$ samples with $x =$ 0.02 (a),
0.05 (b), and 0.075 (c), after annealing in oxygen at normal
pressure (closed symbols) and high pressure (open symbols, HPA).
As a result of HPA, the superconducting $T_c$ remains constant,
decreases or increases, depending on the substitution level.}
\label{R(T)}
\end{figure}

Figure~\ref{R(T)} shows normalized resistance for three
YBa$_2$Cu$_{3-x}$Mo$_x$O$_{7+d}$ samples with $x = 0.02$, 0.05,
and 0.075, after NPA and HPA. Clearly, $T_c$ depends on both the
Mo substitution level $x$ and the oxygen content. The NPA samples
show rapid decrease of $T_c$ with $x$ (see also
Fig.~\ref{x(Mo)y(Sr)}) resulting in a transition temperature of
86~K for $x =$ 0.075. The samples with $x =$ 0, 0.02, and 0.05
behave as optimally doped or slightly over-doped; their $T_c$
decreases after the HPA process, which increases the oxygen
content and thus, most likely, enhances the charge-carrier
concentration. The samples with $x > 0.05$ show under-doped
properties; after the HPA process, their $T_c$ increases. Similar
features are also observed for the
YBa$_{2-y}$Sr$_y$Cu$_{3-x}$Mo$_x$O$_{7+d}$ samples with the
Sr-substitution level $y =$ 0.1, 0.5, and 1 (see
Fig.~\ref{x(Mo)y(Sr)}). The normal-state resistivity of
YBa$_2$Cu$_{3-x}$Mo$_x$O$_{7+d}$ displays a non-linear temperature
dependence that changes with both $x$ and $d$. This behavior, as
well as complex dependence of the superconducting transition
width, cannot be easily interpreted for polycrystalline materials;
we investigate these effects by studying the transport properties
of the recently synthesized single crystals.

The shapes and sizes of the grains were determined by SEM. Most of
the grains have a plate-like shape, are well connected, and have a
size distribution ranging from 5 to 20~$\mu$m, with an average
main plate dimension of about 10 and 8~$\mu$m for the samples with
$x \leq 0.02$ and $x \geq 0.05$, respectively. Thus, to estimate
the effective superconducting volume fraction of the grains and to
evaluate the density of the intragrain persistent critical
currents, an average grain size of 10 and 8~$\mu$m was taken for
the samples with $x \leq 0.02$ and $x \geq 0.05$, respectively.
Adopting the main plate dimension as a characteristic scaling
length for the randomly oriented grains results in a certain
underestimation of the intragrain critical currents calculated
from the Bean model. Note, that for the YBaSrCu$_3$O$_{7+d}$
compound, studied previously, the grains were smaller (3-4~$\mu$m)
and more round revealing a systematic decrease of the grain size
with Sr-substitution in the Y123 compound.\cite{Rogacki'PRB'2000}

\section{CRYSTAL STRUCTURE}

Neutron powder diffraction experiments were carried out to
determine the exact structural details of our Mo-substituted YBCO
materials.  Initial refinements were performed assuming the
structural and atomic positions of the parent YBCO material.
Attempts to refine the site occupancies of the two independent
copper sites [chain (Cu1) and planar (Cu2) sites] resulted in the
planar Cu2 sites being completely occupied only by Cu while the
occupancy of the Cu1 chain sites was significantly reduced because
of the Mo substitution on this site.  However, determining of the
exact Cu and Mo site occupancies was not possible because of the
low contrast between their neutron scattering lengths (7.718~fm
and 6.715~fm for Cu and Mo, respectively).  As such, in all
subsequent refinements we fixed the Cu and Mo contents at their
nominal values.  On the other hand, refinements of the oxygen site
occupancies demonstrated that all sites are essentially full
within one or two standard deviation except for the oxygen site
(O1) that is located in the blocking layer between the Cu1 chain
atoms.  Additionally, a significantly larger than expected
temperature factor was refined for this oxygen atom indicating
that atoms occupying this site must be shifted off their special
positions (0 $\frac{1}{2}$ 0).  Subsequent refinements
demonstrated that O1 oxygen atoms actually occupy atomic positions
(x $\frac{1}{2}$ 0) that are slightly shifted off the a-axis and
that their site occupancy decreases with increasing Mo content.  A
second oxygen atom (O5) was introduced at ($\frac{1}{2}$ y 0) and
found to have a site occupancy that increases with increasing Mo
content. As such, the total oxygen content in this layer would be
the sum of both the site occupancies of O1 and O5.  A summary of
all significant structural parameters is listed in
Table~\ref{Table1}.

\begin{table}[!htb]
\includegraphics*[width=0.5\textwidth]{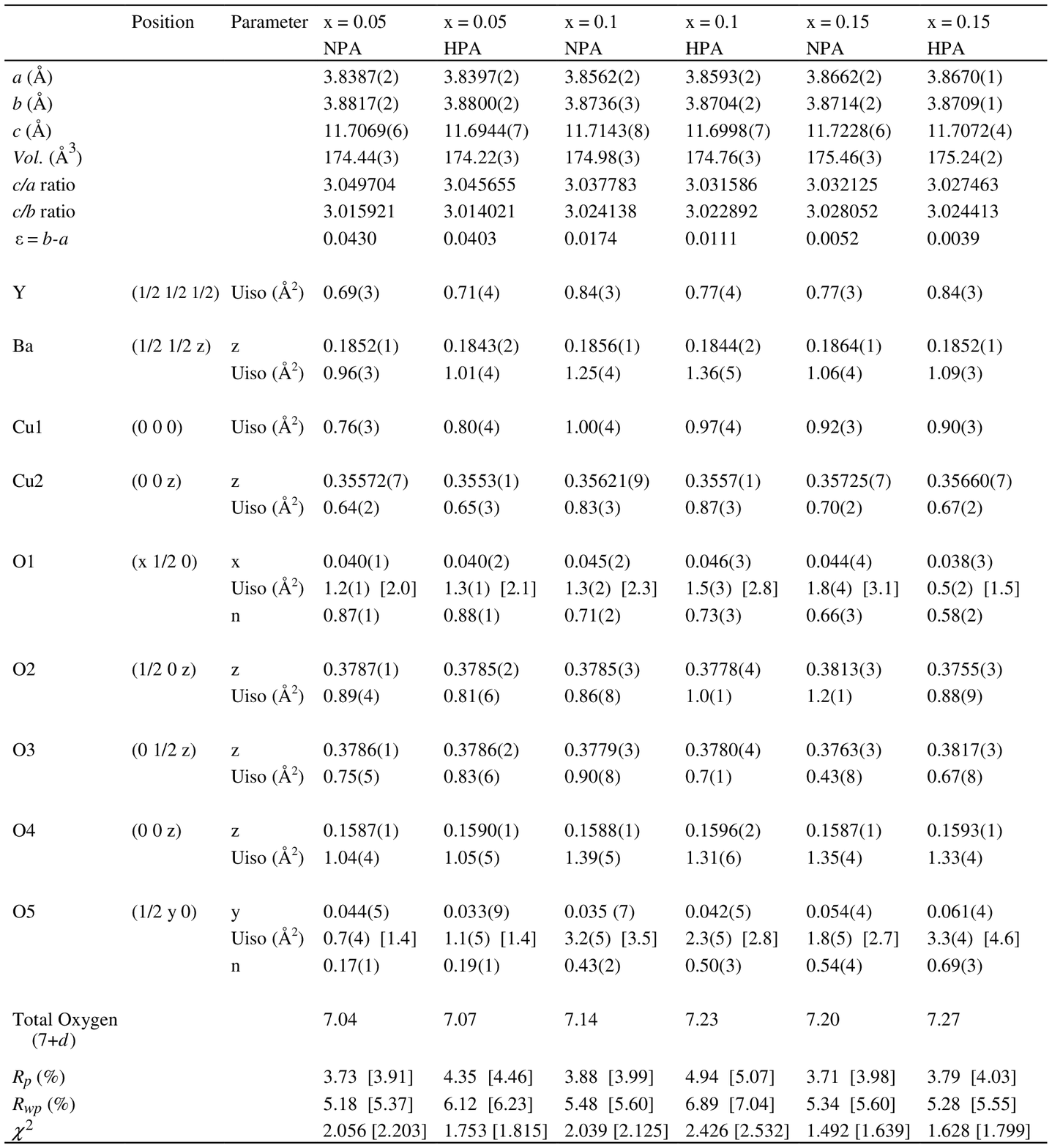}
\caption{Refined structural parameters for
YBa$_2$Cu$_{3-x}$Mo$_x$O$_{7+d}$ samples annealed in oxygen at
normal pressure (NPA) and high pressure (HPA). Values in square
parentheses are obtained from refinements with O1 and O5 occupying
special positions (0 $\frac{1}{2}$ 0) and ($\frac{1}{2}$ 0 0),
respectively.} \label{Table1}
\end{table}

The careful examination of the structural results leads to several
important conclusions: 1- the orthorhombicity factor ($\epsilon =$
b-a) of the samples decreases significantly as the Mo content
increases. Notice that the a-axis increases while the b-axis
decreases as a function of increasing Mo content.  2- the site
occupancy of O1 decreases significantly from 0.87 for $x =$ 0.05
to 0.66 for $x =$ 0.15 for the as-made samples and that the
high-pressure annealing either leaves the occupancy of this site
unchanged or reduces its content.  3- the site occupancy of O5
increases significantly from 0.17 to 0.43 and 0.54 for the NPA $x
=$ 0.05, 0.1 and 0.15 samples, respectively.  High-pressure
annealing results in increasing the oxygen content of this site
contrary to what is observed for the O1 site.  4- for the NPA
samples, the overall oxygen content increases up to 7.20 atom per
formula unit with increasing Mo content.  5- the high-pressure
annealing forces extra oxygen atoms into the structure with the
oxygen content increasing from 7.04 to 7.07, 7.14 to 7.23, and
7.20 to 7.27 for the $x =$ 0.05, 0.1 and 0.15 samples,
respectively. Here we note that the Mo-substituted samples
prepared under normal oxygen pressures possess oxygen contents
that are significantly greater than the oxygen content of the
unsubstituted YBCO parent material ($\simeq 6.97$) (see the total
oxygen contents near the bottom of Table~\ref{Table1}).  Thus, the
difference in oxygen content between the substituted and
unsubstituted materials may be directly attributed to the presence
of Mo atoms in the host structure.  As such, the $x =$ 0.05, 0.1
and 0.15 samples exhibit additional 0.07(2), 0.17(2) and 0.23(2)
oxygen atoms, respectively (assuming an oxygen content of 6.97 for
the parent YBCO material).  We can easily note that the ratio
between the extra oxygen atoms and Mo content is nearly 3:2 for
all the samples, which means that three additional oxygen atoms
are introduced into the structure for every pair of Mo atoms. This
ratio can be explained by the presence of Mo pairs pointing
preferentially in the direction of the a-axis in a fashion similar
to that observed for YBaSrCu$_{3-x}$Mo$_x$O$_{7+d}$ and
YSr$_2$Cu$_{3-x}M_x$O$_{7+d}$ ($M$ =
Mo,~W).\cite{Rogacki'PRB'2000,Dabrowski'PhysicaC'1997} We note
that the other ordering schemes are also possible to preserve the
2:3 ratio of the added Mo and oxygen atoms, respectively. For
example, four MoO$_6$ octahedra replacing a square of four Cu
atoms in the chain layers would give the same ratio. Larger
clusters are also possible with the same ratio. However, as the
sizes of clusters grow, we would need to consider them as
inclusions of an impurity phase, MoBa$_2$YCu$_2$O$_8$, which we do
not observe in our diffraction measurements. Thus, the
Mo$_2$O$_{11}$ and possibly the limited amount of Mo$_4$O$_{20}$
clusters seem to be the most likely defects explaining
consistently our results. The clusters such as dimers or squares
of MoO$_6$ octahedra in the CuO chains disturb locally the
crystallographic structure and the oxidation state of the Cu
atoms, and interfere with the charge transfer between the chains
and planes, and thus perturb locally the superconducting state.
These effects, which may result in creation of pinning centers,
are discussed in the following chapter.

\section{MAGNETIC CHARACTERIZATION AND CRITICAL CURRENTS}

\subsection{Superconducting Volume Fraction}

The shielding effect and the effective superconducting volume
fraction of grains were estimated from the ac susceptibility
measurements performed at 5~K.  The average density of the solid
samples was measured to be 5.1~($\pm 0.1$)~g/cm$^3$; i.e. about
80~\% of the theoretical material density (6.3 g/cm$^3$). The
absolute $\chi'$ values for solid samples were obtained using the
measured sample volume and correcting the data by taking into
account demagnetizing effects. For our solid samples, for all
compositions, the corrected $\chi'$ values are about 5~\% lower
than the ideal value, -1/4$\pi$, which would indicate perfect
shielding. The effective superconducting volume fraction of the
grains can be, in principle, estimated for both solid and powder
samples when the grains are decoupled in magnetic fields large
enough to depress superconductivity in the intergrain region. For
our solid samples, it was not possible to separate grains even at
quite high fields and temperatures. Thus, powder samples were used
for evaluation of the superconducting volume fraction, critical
currents, and the irreversibility field. For powder samples,
almost complete grain separation is observed for a field of 1~T at
all temperatures down to 5~K. Taking into consideration the grain
demagnetizing factor ($N \approx 0.33$), the effective
superconducting volume fraction of grains is estimated to be
between 35 and 65 \%, depending on the substitution level. This
confirms bulk superconductivity and, moreover, attests to good
quality of our samples when compared with Y123 material ($\sim 70$
\%). \cite{Chen'PhysicaC'1990,Campbell'Cryogen'1991}

\subsection{Critical Current Density}

\begin{figure}[!htb]
\includegraphics*[width=0.35\textwidth]{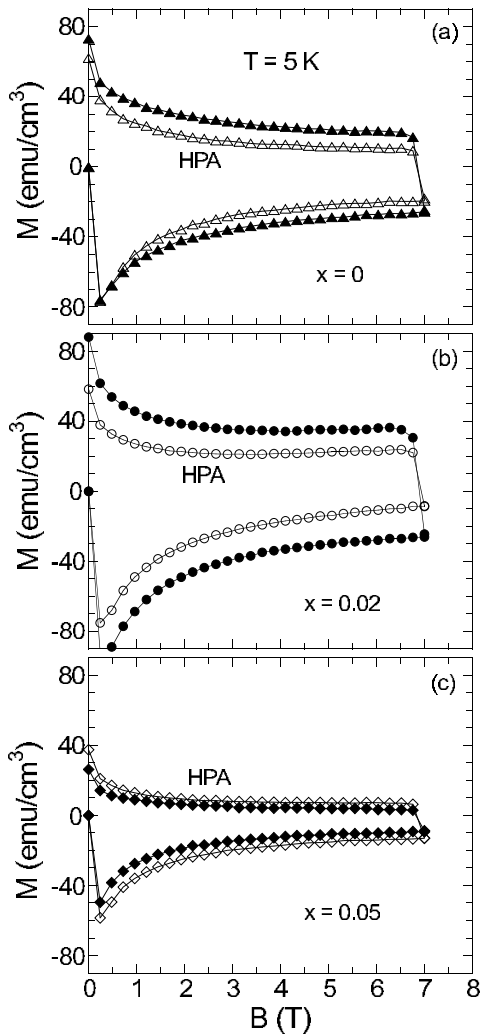}
\caption{Magnetization half-loops measured at 5 K for the
YBa$_2$Cu$_{3-x}$Mo$_x$O$_{7+d}$ samples with $x =$ 0 (a), 0.02
(b), and 0.05 (c) annealed in oxygen at normal pressure (closed
symbols) and high pressure (open symbols, HPA).} \label{M(B)5K}
\end{figure}

\begin{figure}[!htb]
\includegraphics*[width=0.34\textwidth]{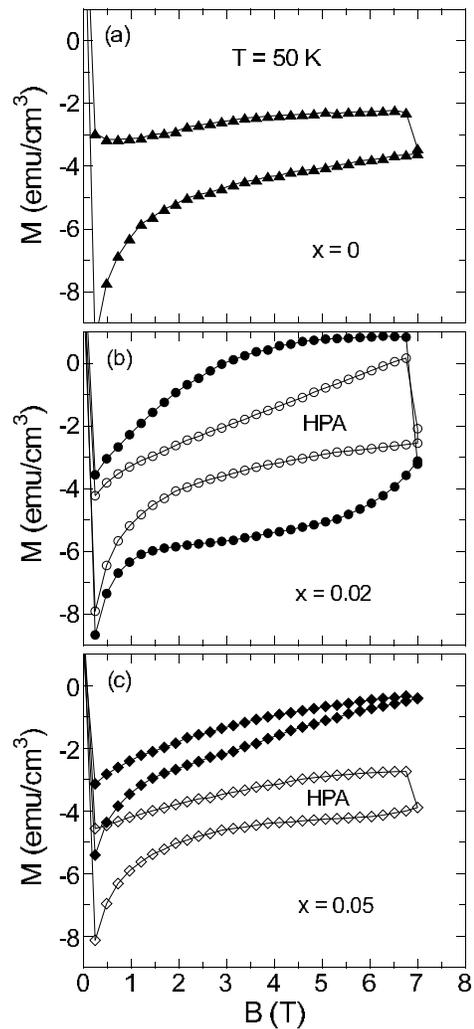}
\caption{Magnetization half-loops measured at 50 K for the
YBa$_2$Cu$_{3-x}$Mo$_x$O$_{7+d}$ samples with $x =$ 0 (a), 0.02
(b), and 0.05 (c) annealed in oxygen at normal pressure (closed
symbols) and high pressure (open symbols, HPA). A large fishtail
effect appears for the Mo-substituted sample with $x =$ 0.02.}
\label{M(B)50K}
\end{figure}

Magnetization hysteresis loops, $M(B)$, were measured at constant
temperatures for powder samples to determine the intragrain
persistent critical current density, $j_c$.  Several examples of
the $M(B)$ curves are shown in Fig.~\ref{M(B)5K} and
Fig.~\ref{M(B)50K} for $T =$ 5 and 50~K, respectively. By
comparing the shapes and sizes of these hysteresis loops, it
becomes immediately clear that the $j_c(B)$ properties change
significantly with the temperature and degree of Mo substitution.
Such behavior is expected when the dominant pinning mechanism
varies between different types, as discussed in
Ref.~\onlinecite{Balents'PRB49'13030'1994}. The irreversibility of
the magnetization, which is proportional to the persistent
critical current density, is the largest for the $x =$ 0.02
sample. In addition, a pronounced widening of the magnetization
loop, the fishtail effect, is observed for that sample at 50~K
(see Fig.~\ref{M(B)50K}). The high oxygen pressure anneal reduces
the irreversibility of the $M(B)$ loops, however, it
simultaneously shifts the maximum of the fishtail effect to much
higher fields. This suggests that HPA creates new types of pinning
centers that are effective at higher vortex densities, but also
suppresses those which are present for samples with smaller oxygen
content. At 5~K (see Fig.~\ref{M(B)5K}), no fishtail effect is
observed for any sample. This may be explained by the fact that at
low temperatures the superconducting coherence length, $\xi$, is
comparable to the distance between the CuO$_2$ double planes,
$\sim 8.5$ \AA, and the region between the double planes acts as
the dominant intrinsic pinning center for every sample studied.
The reduction of the $M(B)$ loops by HPA for $x =$ 0 and 0.02 may
be caused in part by a small decrease in $T_c$ (see
Fig.~\ref{R(T)}). The $x =$ 0.05 sample displays a noticeable
increase of the $M(B)$ loops after HPA which does not change
$T_c$. One reason for this increase may be additional oxygen,
introduced into the chains region, which can act as another weak,
point-like pinning centers. This is supported by structural
results obtained from neutron powder diffraction and by
thermogravimetric measurements which show that the HPA increases
the oxygen content from $d =$ 0.04 to 0.07. The $M(B)$ results
show that the optimal Mo substitution level for the largest $j_c$
is around $x =$ 0.02.

The extended Bean critical-state model was applied to calculate
the persistent critical current density from magnetization
measurements by using the relation $j_c$(A/cm$^2$)~= $k\Delta
M/w$, where $\Delta M$ is the width of the $M(B)$ loop in
emu/cm$^3$, $w$ is a scaling length in cm, and $k$ is a shape
coefficient.\cite{Civale'PRB'1991} For our randomly oriented
plate-like grains, $k =$ 30 was taken judicially. The $\Delta
M(B)$ values at fields above the first magnetization peak were
used to make the extended Bean model
applicable.\cite{Kumar'PRB'1989} These fields are close to or
above the full penetration filed, so they approximately satisfy
the condition of the parallel orientation of the external field.
Consequently, the component of $j_c$ parallel to the external
field seems to be negligible compared to the component
perpendicular to the field, and thus, it is possible to accurately
determine the value of $j_c$. The smallest values of $\Delta M$
which could be reliably separated from the noise of the
magnetometer were approximately $1 \cdot 10^{-4}$ and $0.5 \cdot
10^{-4}$~emu/g (corresponding to the critical current densities of
about 20 and 10~A/cm$^2$) for the pure and substituted samples,
respectively. The grain diameter was taken as the size of the
current loops, $w$, for dc applied fields high enough to separate
grains. This was deduced from the ac susceptibility measurements
performed at various dc fields.  The separation of grains was
confirmed by the absence of a second maximum in $\chi''(T)$ that
would mark the transition to the superconducting state for the
intergrain material.  The grain decoupling field was about 1~T at
temperatures of about 60, 55 and 50~K for the powder samples with
$x =$ 0, 0.05 and 0.1, respectively. However, for fields above
1~T, the intergrain coupling was estimated to be small even at
5~K. Thus, we conclude that for fields above 1~T, the intergrain
component of $j_c$ can be neglected at all relevant temperatures
and the average grain diameter can be taken as the scaling length.

\begin{figure}[!htb]
\includegraphics*[width=0.38\textwidth]{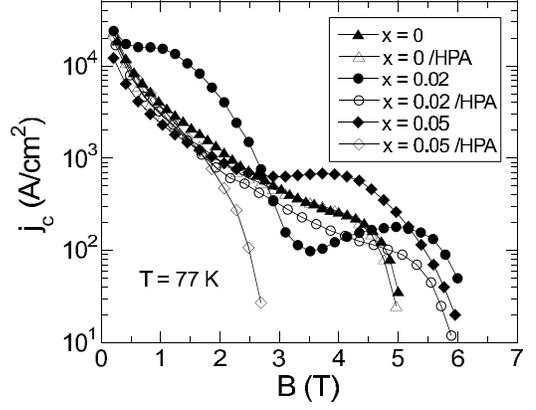}
\caption{Persistent critical currents, $j_c$, versus magnetic
field at 77 K for the YBa$_2$Cu$_{3-x}$Mo$_x$O$_{7+d}$ samples
with $x =$ 0 (triangles), 0.02 (circles), and 0.05 (diamonds)
annealed in oxygen at normal pressure (closed symbols) and high
pressure (open symbols, HPA).} \label{j(B)77K}
\end{figure}

Figure~\ref{j(B)5,35,50K} shows $j_c$ as a function of magnetic
field at temperatures of 5, 35, and 50~K for the
YBa$_2$Cu$_{3-x}$Mo$_x$O$_{7+d}$ samples with $x =$ 0, 0.02, 0.05,
and 0.075 prepared at NPA and HPA conditions. At 5 K, the nearly
field-independent $j_c$ is the largest for $x =$ 0.02 at about $2
\cdot 10^6$~A/cm$^2$. At this temperature, the difference between
the $x =$ 0 and 0.02 samples is relatively small, most likely
because the coherence length, $\xi$, is similar to the width of
the pinning region between the superconducting CuO$_2$ double
planes. The influence of the Mo substitution is clearly observed
at elevated temperatures, where the $x =$ 0.02 sample shows a huge
increase of $j_c$ over the $x =$ 0 sample. While enhancement as
large as 30 to 50~\% is observed at low temperatures, the
substitution-created peak effect (PE) leads to 100 and 250 \%
increase of $j_c$ at 35 and 50~K, respectively.
Figure~\ref{j(B)77K} shows $j_c(B)$ at 77~K for the $x =$ 0, 0.02,
and 0.05 samples prepared at NPA and HPA conditions. In a field of
1.5~T, $j_c$(1.5 T) $\simeq 2\cdot 10^3$~A/cm$^2$ for the
non-substituted sample and increases to about $1 \cdot
10^4$~A/cm$^2$ for the substituted sample with $x =$ 0.02. At
higher fields, $j_c$ falls below 100~A/cm$^2$ at 5 and 6~T for the
$x =$ 0 and 0.02 samples, respectively. The HPA process affects
$j_c$ in two ways. For the pure and slightly substituted samples,
HPA barely reduces $T_c$ while strongly suppressing $j_c$. This is
most likely because by introducing a small amount of excess oxygen
into the chain region, the HPA produces disordered point-like
defects that reduce the superconducting volume fraction of the
material. However, at moderate temperatures, HPA shifts PE to
higher fields (see Fig.~\ref{j(B)5,35,50K}) probably as a result
of weakly interacting pinning centers originated from these
point-like defects. For samples with a higher substitution level
$(0.05 \leq x \leq 0.075)$, the HPA increases $j_c$ in the wide
temperature range from 5 to 70 K. This increase can be
qualitatively explained as a result of charge doping that
increases $T_c$ for $x \geq 0.05$. At temperatures close to $T_c$,
HPA decreases $j_c$ for all substituted samples, most likely due
to the point-like defects that can reduce the irreversibility
field, $B_{irr}$.\cite{Wolf'PRB'1997} A significant reduction of
$B_{irr}$ by HPA has been observed for our substituted samples;
this effect will be discussed in the next section.

\begin{figure}[!htb]
\includegraphics*[width=0.38\textwidth]{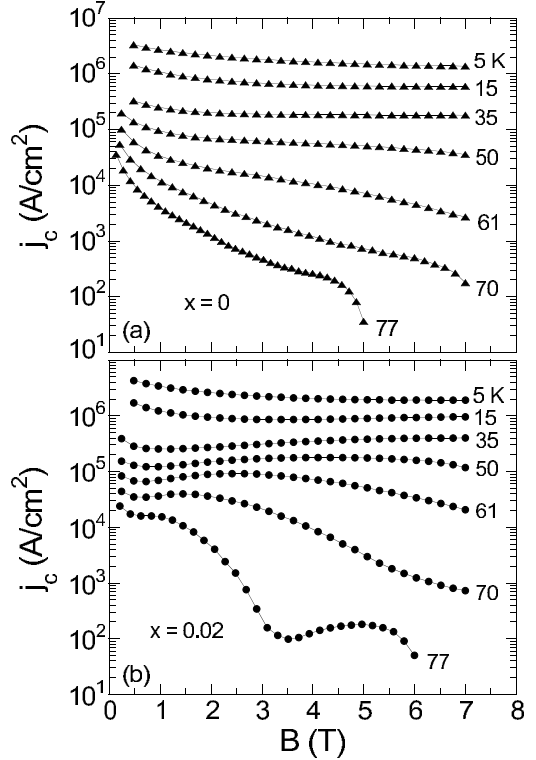}
\caption{Persistent critical currents, $j_c$, as a function of
magnetic field at several temperatures from 5 to 77 K for the
YBa$_2$Cu$_{3-x}$Mo$_x$O$_{7+d}$ samples with $x =$ 0 (a) and 0.02
(b) annealed in oxygen at normal pressure.} \label{j(B)5-77K}
\end{figure}

Comparison of critical current properties at several temperatures
between 5 and 77~K is shown in Fig.~\ref{j(B)5-77K} for NPA
samples of YBa$_2$Cu$_{3-x}$Mo$_x$O$_{7+d}$ with $x =$ 0 and 0.02.
There are three regions of $j_c$ dependence on $B$ for $x =$ 0. At
low fields, $j_c$ has a weaker than exponential decrease for all
temperatures, similar to other high temperature superconductors.
At higher fields below 35~K, a roughly exponential dependence is
observed, extending to at least 7~T. At higher temperatures, the
range of exponential dependence narrows and is followed by a more
rapid decrease, especially above 70~K. Thus, PE is not observed
for the $x =$ 0 sample. Figure~\ref{j(B)5-77K}(b) shows that for
$x =$ 0.02 the region of the exponential decrease of $j_c(B)$
exists only at temperatures below 15~K.  At higher temperatures, a
large maximum appears, shifting to lower fields with increasing
temperature. This PE is a result of the fishtail effect seen for
the magnetization loops (Fig.~\ref{M(B)50K}). For temperatures
above 65~K, the second maximum of $j_c(B)$ develops at high
fields. The increase of $j_c$ for $x =$ 0.02 seems to be more
remarkable at 70~K, where the second maximum develops at fields
higher than 7~T. At low fields and in the single-vortex pinning
regime, a simple summation of the individual microscopic pinning
forces leads to the relation $j_c \sim B^\alpha$, where $\alpha =$
-1/2.\cite{Murakami'SST'1991,Murakami'Cryogen'1992} For our
samples with $x =$ 0, 0.05, and 0.075, this relation can be
observed at fields $B < 1.5$~T in the temperature range from 5 to
65~K. However, for the samples with $x =$ 0.05 and 0.075, $j_c
\sim B^\alpha$ with $\alpha =$ -1/2 is always observed over a
narrower magnetic field range when compared with the sample with
$x =$ 0. For the optimally substituted compound with $x =$ 0.02,
no such dependence of $j_c(B)$ is observed. These results suggest
that collective rather than single-vortex pinning is present at
moderate fields for the Mo-substituted samples. At higher fields,
the characteristic $j_c(B) \sim B^\alpha$ dependence has not been
observed over any extended range of field or temperature making
the analysis of the pinning mechanism more difficult and less
useful.

Figures~\ref{j(B)5,35,50K} through \ref{j(B)5-77K} show that the
significant effect of the Mo-substitution on the pinning
properties of Y123 is present at elevated temperatures and fields
where a large PE in $j_c(B)$ is observed. At elevated
temperatures, point-like defects such as individual Mo-ions in the
CuO chains are not capable of effective pinning because the
interaction potential is smeared out by the thermal oscillation of
vortices.\cite{Wolf'PRB'1997,Kupfer'PRB'1996} Thus, we suggest
that the PE observed in our Mo-substituted samples at temperatures
above 30~K is the result of more extended defects. This is
consistent with our neutron diffraction results and oxygen-content
measurements which provide evidence for existence of such extended
defects in the form of the Mo$_2$O$_{11}$ dimers (and possibly
also Mo$_4$O$_{20}$ clusters) formed from nearest neighbor MoO$_6$
octahedra. This conclusion is in agreement with results reported
for the melt textured Nd123 compound, where the pronounced PE was
observed and interpreted as a result of weakly-interacting, small
clusters of point-like defects.\cite{Wolf'PRB'1997}

\begin{figure}[!htb]
\includegraphics*[width=0.38\textwidth]{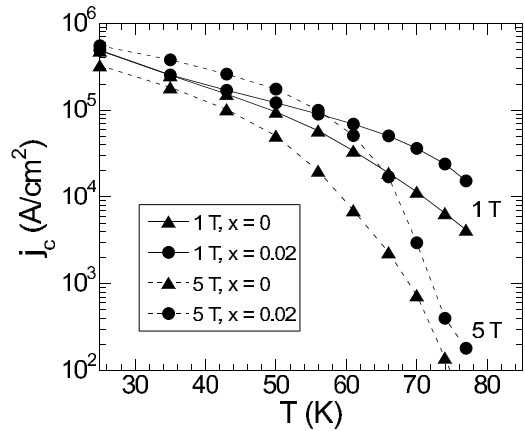}
\caption{Temperature dependence of persistent critical currents,
$j_c$, for the YBa$_2$Cu$_{3-x}$Mo$_x$O$_{7+d}$ samples with $x =$
0 (triangles) and 0.02 (circles) annealed in oxygen at normal
pressure. The $j_{c}(T)$ dependencies are shown at the fields of 1
T (solid line) and 5 T (broken line).} \label{j(T)}
\end{figure}

Superior pinning properties of the $x =$ 0.02 sample are more
obvious when $j_c$ is plotted as a function of temperature.
Figure~\ref{j(T)} shows $j_c(T)$ for the $x =$ 0 and 0.02 samples
at fields of 1 and 5~T.  The $j_c(T)$  shows negative curvature
for both $x =$ 0 and 0.02 from 30 to 70~K in high fields. However
at 1~T, whereas $j_c(T)$ for the $x =$ 0 sample shows the same
negative curvature, $j_c(T)$ for the $x =$ 0.02 sample shows
positive curvature below 50~K. The resulting increase of $j_c(T)$
at 1~T over the values obtained for the $x =$ 0 sample can be well
modeled for temperatures between 50 and 80~K by an exponential
expression $j_c(T) = j_c(0)$exp$[-3(T/T^\ast)^2]$ with two fitting
parameters: $j_c(0) = 4.3 \cdot 10^5$~A/cm$^2$ and $T^\ast =
0.85Tc$. This expression was proposed to account for pinning by
correlated disorder\cite{Nelson'PRL'1992,Nelson'PRB'1993} and was
observed for Bi$_2$Sr$_2$CaCu$_2$O$_8$ single crystals with
columnar defects generated by heavy ion
irradiation.\cite{Moshchalkov'PRB'1994} This similarity suggests
that the observed increase of $j_c$ for Mo-substituted samples
could be caused by pinning centers that are partially ordered
along the $c$-axis. An important point to note is that at 1~T the
$j_c(T) \gtrsim 10^4$~A/cm$^2$ shows a maximal enhancement at
about 72-76~K, i.e., at temperatures close to the liquid nitrogen
temperature. These values of $j_c$, observed for the $x =$ 0.02
polycrystalline sample, are similar to, or higher than, those
obtained for the textured Y123/Y211
composites.\cite{Martinez'PRB'1996}  Because in polycrystalline
samples $j_c$ is determined by the current component perpendicular
to the $ab$-plane, even higher critical currents are expected for
the Mo-substituted textured materials when $j_c$ flows in the
$ab$-plane only.

\subsection{Irreversibility line and pinning analysis}

The irreversibility field, $B_{irr}$, was derived from the ac
susceptibility measured as a function of temperature at constant
applied dc fields. Each "irreversibility point"
$(B_{irr},T_{irr})$ of the irreversibility line $B_{irr}(T)$ was
obtained by determining a characteristic temperature, $T_{irr}$,
at which the imaginary part of the ac susceptibility, $\chi''$,
begins to differ from zero.  Below this temperature, both the
critical current density and the ac losses differ from zero; i.e.,
the magnetization is irreversible. Thus, $T_{irr}$ denotes a
temperature below which resistivity is zero and a diamagnetic
signal appears for the real part of the ac susceptibility,
$\chi'$.\cite{Civale'PRB'1991} Because the sensitivity of our ac
susceptibility measurements is at least 100 times higher than that
of our dc magnetization measurements ($2 \cdot 10^{-5}$~emu), the
$B_{irr}(T)$ curves obtained from ac susceptibility are taken as
the upper limits of the $B_{irr}(T)$ lines derived from
magnetization.\cite{Rogacki'PRB'2000}

\begin{figure}[!htb]
\includegraphics*[width=0.38\textwidth]{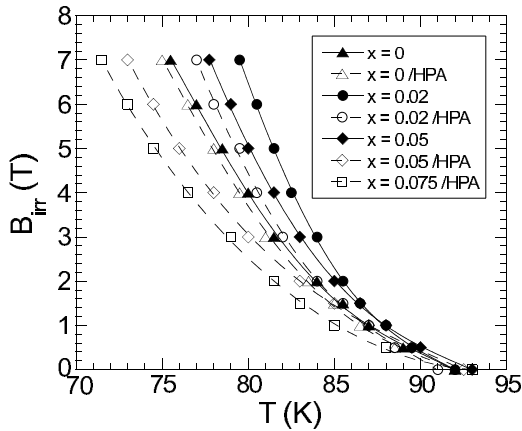}
\caption{Temperature dependence of the irreversibility field,
$B_{irr}$, obtained from ac susceptibility measurements for the
YBa$_2$Cu$_{3-x}$Mo$_x$O$_{7+d}$ samples with $x =$ 0 (triangles),
0.02 (circles), 0.05 (diamonds), and 0.075 (squares) annealed in
oxygen at normal pressure (closed symbols) and high pressure (open
symbols, HPA). The lines are guides for the eye only.}
\label{Birr(T)}
\end{figure}

The irreversibility lines determined by ac susceptibility are
shown in Fig.~\ref{Birr(T)} for the NPA and HPA powder samples.
The highest irreversibility line and the largest slope of
d$B_{irr}$/d$T$ were obtained for the NPA sample with $x =$ 0.02.
Increasing the Mo-substitution shifts the $B_{irr}(T)$ line to
lower fields. However, even for the $x =$ 0.05 sample, the line is
still above that of the $x =$ 0 sample.  A large decrease of
$B_{irr}(T)$ is observed for all Mo-substituted samples after HPA.
This decrease may be caused by introduction of uncorrelated
defects (additional oxygen ions) and is consistent with results
obtained for $j_c(B)$ closer to $T_c$ (see Fig.~\ref{j(B)77K})
where a sudden drop of $j_c$ below 10~A/cm$^2$ appears at fields
that correspond to $B_{irr}$. A small difference between $B_{irr}$
and fields where the drop appears is caused by different
sensitivity in measurements used to obtain $B_{irr}$ (ac
susceptibility) and $j_c$ (dc magnetization). The temperature
dependence of $B_{irr}(T)$ can be expressed by a power-law
relation $B_{irr} \sim (1 - T/T_c)^n$, with $n \simeq$ 1.7 and 1.6
derived for the $x =$ 0 and 0.02 samples, respectively. For $x =$
0, the power relation with $n =$ 1.7 fits very well over the whole
temperature range shown from 75~K to $T_c =$ 92~K. For $x =$ 0.02,
the power law holds only over much narrower temperature range from
about 83~K to $T_c$. A small difference of the exponent near $T_c$
points to rather minor changes of the electronic structure
introduced by the Mo substitution. Strongly divergent behavior of
the $B_{irr}(T)$ at lower temperatures for the $x =$ 0.02 sample
results from increased volume pinning force.

From the above-analyzed behavior of $j_c$, $B_{irr}$, and $T_c$,
it appears that the Mo-substitution mainly changes the pinning
properties of the Y123 material. The substitution introduces
pinning centers that are most likely Mo$_2$O$_{11}$ dimers located
in the CuO chains, as concluded from the neutron results. The
dimers may act as pinning centers because they induce extended
distortions in the structure and, therefore, may locally perturb
superconductivity in the CuO$_2$ planes.  The individual
perturbation with a dimension of 2-3 unit cells in the $ab$-plane
is comparable to the superconducting coherence length, $\xi$, and
therefore may act as an effective pinning center.  It is plausible
that the dimers order partially along the $c$-axis to form a
number of oriented "columnar"
defects.\cite{Chmaissem'PRB'1995,Jorgensen'LNP'1996} To analyze
the relevant pinning mechanism related to the pinning centers
introduced by the Mo-substitution, we turn now to discussion of
the volume pinning force as a function of field by using the
scaling procedure.

The volume pinning force is defined by a formula $F_p(B) = j_c
\cdot B$, where $F_p$ is in N/m$^3$, $j_c$ in A/m$^2$, and $B$ in
T (i.e., the main features of the $F_p(B$) are similar to $j_c(B)$
as shown in Figs.~\ref{j(B)5,35,50K}-\ref{j(B)5-77K}). Two
maximums in $j_c(B)$ and $F_p(B)$ are observed at 77~K for the NPA
sample with $x =$ 0.02 (see Fig.~\ref{j(B)77K}). The first maximum
appears at low fields ($B \simeq 1.2$~T) and the second at higher
fields ($\simeq 5$~T) close to a critical field where a sudden
drop of the pinning force is observed. Such behavior of $F_p(B)$
is generally consistent with the prediction of the static
collective pinning theory for the pinning centers of average
force;\cite{Larkin'JLTP'1979} for example, small clusters of
dislocations like the Mo$_2$O$_{11}$ dimers. For the sample with a
larger Mo-substitution level ($x =$ 0.05, see Fig.~\ref{j(B)77K}),
the low-field maximum disappears while the high-field maximum
increases above that observed for the $x =$ 0.02 sample. According
to the static collective pinning theory, this behavior indicates
that the pinning arises from weakly interacting centers as a
function of increased concentration. This is in fact the case for
our highly substituted samples, where an introduction of
additional Mo$_2$O$_{11}$ dimers results in a denser and therefore
smoother pinning structure. Such a pinning structure is
characterized by a distance between pinning centers that is only
slightly larger than $\xi$; i.e., the pinning matrix is expected
to have weaker interaction with vortices. Using this model, the
observed elimination of both maximums in $F_p(B)$ due to the HPA
process (see Fig.~\ref{j(B)77K}) can be explained in a similar
way. By introducing additional oxygen into the CuO chains, the HPA
creates additional pinning centers, and, in this manner, makes the
pinning structure more uniform.  This explanation is supported by
an observation of increase in $j_c$ at lower temperatures (e.g. at
50~K, see Fig.~\ref{j(B)5,35,50K}) where $\xi(T)$ decreases and
the pinning structure becomes more granular. With decreasing
temperature, an increase of the volume pinning force is expected
to shift the maximum of $F_p(B)$, observed for the NPA sample, to
higher fields. This shift is clearly visible for the $x =$ 0.02
sample (see Fig.~\ref{j(B)5-77K}b). In addition, the second
maximum in $F_p(B)$, that is absent for the $x =$ 0 sample, seems
to shift above 7~T at 70~K. The drastic drop of the pinning force
at these fields and temperature for the $x =$ 0 sample points
again to a large difference in pinning properties between the $x
=$ 0 and 0.02 samples.

\begin{figure}[!htb]
\includegraphics*[width=0.37\textwidth]{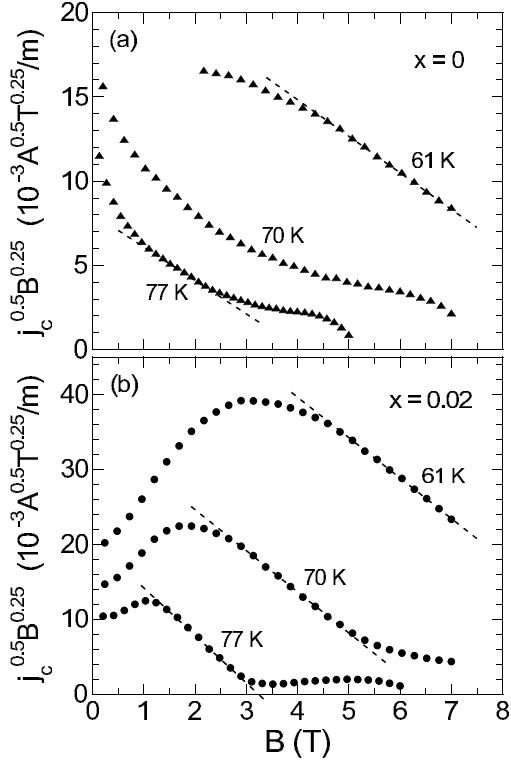}
\caption{Kramer's plots at 61, 70, and 77 K for the
YBa$_2$Cu$_{3-x}$Mo$_x$O$_{7+d}$ samples with $x =$ 0 (a) and 0.02
(b) annealed in oxygen at normal pressure.  Scaling field,
$B_{k}(T)$, is obtained at individual temperature as a field
determined by linear extrapolation of the broken line to $j_c =$
0.} \label{Kramer's}
\end{figure}

When thermal activation effects are negligible, the volume pinning
force is usually described by the following
formula:\cite{Campbell'Adv.Phys'1972,Dew-Hughes'Phil.Mag'1974}
$F_p(b,T) = F_{po}(T)b^p(1-b)^q$, where $F_{po}(T)$ is the
field-independent pinning force in the absence of thermal
activation and $b = B/B^\ast(T)$ is a reduced field, where
$B^\ast(T)$ is the scaling field, and $p$ and $q$ depend on the
relevant pinning mechanism. In this model it is usually difficult
to attribute the involved coefficients to the particular pinning
mechanism, because additional requirements concerning, for
example, the sample geometry and grain orientation have to be
fulfilled. In the Kramer's model of pinning, a simplified scaling
law is used with $f(b) = b^{1/2}(1-b)^2$, where $f =
F_p(b,T)/F_{p,max}(T)$, and $F_{p,max}(T)$ is a maximum volume
pinning force for $F_p(B)$ at each
temperature.\cite{Kramer'J.Apl.Phys'1973} Using Kramer's approach,
$B_{c2}(T)$ is originally taken as the scaling field $B^\ast(T)$.
However, other fields related to $B_{irr}$ rather than $B_{c2}$
have frequently been used to model the scaling
law.\cite{Martinez'PRB'1996,Umezawa'Nature'1993,Higuchi'PRB'1999,
Larbalestier'Nature'2001} Figure~\ref{Kramer's} shows scaled
$j_c^{1/2}B^{1/4}$ versus $B$ (the so-called Kramer's plots),
adopted to obtain the scaling field $B_k(T) = B(T,j_c$=0) by
extrapolating $j_c(B)$ to zero
current.\cite{Larbalestier'Nature'2001,Dew-Hughes'Phil.Mag'1987}
Because $B_k$ obtained this way is a lower limit of the measured
$B_{irr}$, it describes properties related to the global critical
current that flows through the entire sample (grain) rather than
the maximal critical current preserved in some local areas. One
advantage of using $B_k$ is that it can be derived at fields much
higher than those used in actual measurements.  In our case, it
was possible to determine $B_k$ for temperatures down to 61~K;
i.e., the temperatures where $B_{irr}$ is too large to be measured
directly.

\begin{figure}[!htb]
\includegraphics*[width=0.41\textwidth]{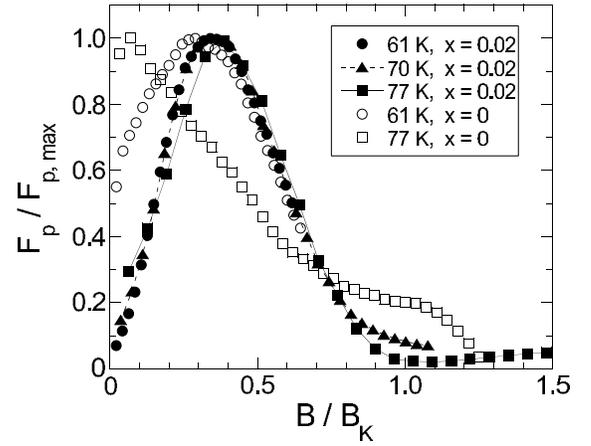}
\caption{Reduced pinning force, $F_{p}/F_{p,max}$, versus reduced
magnetic filed, $B/B_{k}$, for the non-substituted ($x =$ 0, open
symbols) and Mo-substituted ($x =$ 0.02, closed symbols)
YBa$_2$Cu$_{3-x}$Mo$_x$O$_{7+d}$ samples at 61 (circles), 70
(triangles), and 77 K (squares). For the definition of $F_{p,max}$
and $B_{k}$ see text.} \label{Fp(B)}
\end{figure}

The scaling behavior of the volume pinning force is shown above
60~K in Fig.~\ref{Fp(B)} for the $x =$ 0 and 0.02 samples. The
scaling is not expected to hold at lower temperatures where
additional pinning by regions between the CuO$_2$ double planes
dominates. The scaled curves in the temperature range from 61 to
77~K collapse to a single curve for $x =$ 0.02, proving that a
single pinning mechanism is responsible for increase of the
critical currents. This excellent scaling of $F_p(B)$ indicates
that the Mo$_2$O$_{11}$ dimers, the main defects introduced by
substitution, create pinning centers that are dominant at these
temperatures. The flux creep that intensifies at higher
temperatures may cause a small deviation from the scaling law, as
observed at 77~K. Generally, such thermally assisted flux motion
leads to a breakdown of the scaling behavior because no clear
separation between effects of the field and temperature is
possible.\cite{Niel'Cryogen'1992} At elevated temperatures, the
thermally assisted flux flow results frequently in a tail at the
highest fields, exactly what is observed for the $x =$ 0.02 sample
at 77~K. For $x =$ 0, no universal scaling is observed, indicating
that several different pinning mechanisms may be dominant,
depending on fields and temperatures.

\section{SUMMARY AND CONCLUSION}

We have shown that, by synthesizing
YBa$_2$Cu$_{3-x}$Mo$_x$O$_{7+d}$ materials in an oxygen
atmosphere, it is possible to drive the substituted Mo ions into
the chain region of the crystal structure. The Mo ions appear to
form extended defects, most likely the Mo$_2$O$_{11}$ dimers
formed from nearest neighbor MoO$_6$ octahedra. High-pressure
oxygen annealing restores $T_c \approx$ 92~K. By studying the
field and temperature dependence of the intragrain critical
currents, we have shown that the Mo substitution introduces an
important contribution to the irreversibility magnetization in
agreement with our previous studies on the tetragonal
YBaSrCu$_{3-x}$Mo$_x$O$_{7+d}$ superconductor with $T_c \approx$
86~K.\cite{Rogacki'PRB'2000} The large enhancement of the
intragrain critical current density has been found for
YBa$_2$Cu$_{3-x}$Mo$_x$O$_{7+d}$ at elevated fields and
temperatures as a distinct peak effect observed in $j_c(B)$ for
compositions with a small Mo-substitution level $x =$ 0.02. Based
on magnetization data, we infer that the Mo$_2$O$_{11}$ dimers
introduce additional pinning that is dominant at temperatures
above 30~K by locally perturbing superconductivity in the CuO$_2$
planes. We have shown that for the Mo-substituted samples, the
pinning force scales with a characteristic field $B_k$ which is
related to the irreversibility field rather than to the upper
critical field. The excellent scaling behaviour that is observed
over a wide field and temperature range for the Mo-substituted
samples provides additional support for the substitution-induced
pinning centers.

For the composition with $x =$ 0.02, the intragrain critical
current density has been increased by a factor of 3 to 4 at
temperatures around 50~K and fields of about 6~T. An additional
improvement of the critical current density could be expected at
lower fields if ordering of the Mo$_2$O$_{11}$ dimers is achieved
along the $c$-axis. The annealing at high oxygen pressure, which
introduces additional oxygen into the structure, reduces the
irreversibility of the $M(B)$ loops and shifts the maximum of the
fishtail effect to much higher fields. This suggests that HPA
creates a new type of pinning centers that are effective at higher
vortex densities. Thus, for the optimized
YBa$_2$Cu$_{3-x}$Mo$_x$O$_{7+d}$, the best intragrain critical
currents should be obtained for material with $x =$ 0.01 - 0.02
and optimized oxygen content prepared under controlled oxygen
pressure. Additional improvements of $j_c$ are possible for the
YBa$_{2-y}$Sr$_y$Cu$_{3-x}$Mo$_x$O$_{7+d}$ materials when combined
effects of the Mo- and Sr- substitutions and optimized oxygen
content are utilized.

\begin{acknowledgments}
Work at ILT$\&$SR was supported by the Polish State Committee for
Scientific Research (KBN) under Project \mbox{No. 3 T10A 001 26}.
Work at NIU was supported by the NSF-DMR-0302617. IPNS at Argonne
National Laboratory is supported by the U.S. Department of Energy,
BES-MS, under Contract No. W-31-109-ENG-38.
\end{acknowledgments}


\end{document}